\documentclass[useAMS,usenatbib]{mn2e}

\usepackage{graphicx} \usepackage{epsf}
\title[Sounding stellar cycles with Kepler -- I. Strategy for selecting 
targets]{Sounding stellar cycles with Kepler -- I. Strategy for selecting 
targets}

\author[C. Karoff et al.]{C.~Karoff$^{1}$, 
T.~S.~Metcalfe$^{2}$, 
W.~J.~Chaplin$^{1}$, 
Y.~Elsworth$^{1}$\thanks{E-mail: y.p.elsworth@bham.ac.uk}, 
H.~Kjeldsen$^{3}$,
\and T.~Arentoft$^{3}$ 
\& D.~Buzasi$^{4}$\\
$^{1}$School of Physics and Astronomy, University of Birmingham, 
Edgbaston, Birmingham B15 2TT, UK\\
$^{2}$High Altitude Observatory and Scientific Computing Division, NCAR, 
PO Box 3000, Boulder, CO 80307, USA\\
$^{3}$Department of Physics and Astronomy, Aarhus University, DK-8000 
Aarhus C, Denmark\\
$^{4}$Eureka Scientific, Inc., 2452 Delmer Street Suite 100, Oakland, CA 
94602, USA
}
\begin{document}

\date{Accepted --. Received --}

\pagerange{\pageref{firstpage}--\pageref{lastpage}} \pubyear{2009}

\maketitle

\label{firstpage}

\begin{abstract}

The long-term monitoring and high photometric precision of the {\it Kepler} 
satellite will provide a unique opportunity to sound the stellar cycles of 
many solar-type stars using asteroseismology. This can be achieved by 
studying periodic changes in the amplitudes and frequencies of the 
oscillation modes observed in these stars. By comparing these measurements 
with conventional ground-based chromospheric activity indices, we can 
improve our understanding of the relationship between chromospheric 
changes and those taking place deep in the interior throughout the stellar 
activity cycle. In addition, asteroseismic measurements of the convection 
zone depth and differential rotation may help us determine whether stellar 
cycles are driven at the top or at the base of the convection zone. In 
this paper, we analyze the precision that will be possible using {\it Kepler} to 
measure stellar cycles, convection zone depths, and differential rotation. 
Based on this analysis, we describe a strategy for selecting specific 
targets to be observed by the {\it Kepler} Asteroseismic Investigation for the 
full length of the mission, to optimize their suitability for probing 
stellar cycles in a wide variety of solar-type stars.

\end{abstract}


\section{Introduction}

Despite the enormous efforts that have been carried out over the past 400 
years to observe and understand the solar cycle, the delayed onset of 
cycle 24 suggests that we still lack a complete understanding of this 
phenomenon \citep{zimmerman}. One of the best ways to improve our understanding may be to study activity cycles in other stars, to help us understand stellar cycles 
in general and not just the solar cycle. Asteroseismology is emerging as 
one of the best tools available for the study of stellar cycles.

Astronomers have been making telescopic observations of sunspots since the 
time of Galileo, gradually building a historical record showing a periodic 
rise and fall in the number of sunspots every 11 years. We now know that 
sunspots are regions with an enhanced local magnetic field, so this 
11-year cycle actually traces a variation in surface magnetism. Attempts 
to understand this behavior theoretically often invoke a combination of 
differential rotation, convection, and meridional flow to modulate the 
field through a magnetic dynamo \citep[e.g.,][]{2006ApJ...649..498D}.

Significant progress in dynamo modeling could only occur after 
helioseismology provided meaningful constraints on the Sun's interior 
structure and dynamics \citep{1989ApJ...343..526B, 1998ApJ...505..390S}. 
Generally the models assume that the driving of the solar dynamo is 
related to large gradients in the rotation rate of different layers of the 
Sun, presumably those layers with the largest gradients, i.e. at the top 
and bottom of the near surface convection zone 
\citep{2005ApJ...625..539B}.

Stellar activity has been monitored in more than 100 stars over the last 
40 years with the Mount Wilson survey \citep{1978ApJ...226..379W, 
1985ARA&A..23..379B, 1995ApJ...438..269B}. This survey has revealed that 
around half of the solar-type stars show clear periodic cycles, with 
periods between 2.5 and 25 years \citep{1995ApJ...438..269B}. Later 
studies of this and other samples such as the Lowell Observatory survey 
\citep{2007AJ....133..862H} suggest that there are two different branches 
of stellar cycles in solar-type stars -- one active and one inactive 
\citep{1999ApJ...524..295S}. This bifurcation, which is also known from 
the Vaughan-Preston gap \citep{1980PASP...92..385V} between active stars 
younger than 2.5 Gyr and older less active stars, can also be seen in the 
lengths of the stellar cycles as a function of the rotation period. 
\citet{2007ApJ...657..486B} suggested, based on the fact that the number 
of rotations during a cycle seems to be different between the two subgroups of stars, that the bifurcation is caused by different dynamos operating in 
the two branches. Active stars generally have around 300 rotations per cycle, 
whereas inactive stars have around 100 rotations per cycle. According to 
\citet{2007ApJ...657..486B}, this suggests that stellar cycles in active 
stars are generated by the gradient in the rotation rate close to the 
surface, whereas the stellar cycles in inactive stars are generated by the 
gradient in the rotation rate at the base of the convection zone. In this 
scenario, a star with a given mass will start its life on the main 
sequence as a fast rotating active star with a relatively thin convection 
zone \citep{2007ApJ...657..486B}. Since the convective turnover time in stars with thinner convection zones is relatively short, these stars will have the largest rotational gradients near the surface, which would thus be where the 
dynamo would be driven in the stars on the active sequence. As the star 
evolves to the inactive sequence its surface rotation rate decreases 
significantly, which might be caused by additional deep mixing in a 
growing convection zone \citep{2007ApJ...657..486B}. This would result in the rotational gradient at the base of the convection zone growing larger than the rotational gradient near the surface, and therefore the dynamo would be driven at the base of the convection zone in the inactive stars. The larger rotational 
gradient would also require fewer rotations for the dynamo to create a 
toroidal magnetic field strong enough to rise to the surface.

Unfortunately, the Sun in many ways seems to fall between the two 
sequences of active and inactive stars -- especially if one looks at the 
length of the stellar cycle as a function of the rotation period 
\citep{2007ApJ...657..486B}. This suggests that the Sun is not a good 
place to study stellar cycles, because two different dynamos might be 
operating in the Sun at the same time, confusing the picture. In other 
words, since the Sun seems to be in the transition from the active to the 
inactive sequence, it might be experiencing two different dynamos; one 
operating at the base of the convection zone and one operating at the top 
-- as opposed to most other solar-type stars which are only experiencing 
one of the two dynamos. Evidence of this scenario can also be found in the 
discussion of whether the solar dynamo operates at the base or at the top 
of the convection zone \citep{2005ApJ...625..539B}.

If we are to improve our understanding of the solar cycle, it seems that 
we need to study stellar cycles. To evaluate the different dynamo models, 
we need asteroseismic measurements to complement the activity measurements 
of stellar cycles -- both in the sense of asteroseismic measurements of 
cycle-induced changes in the oscillation mode frequencies and amplitudes, 
and in the sense of asteroseismic measurements of the convection zone 
depth and differential rotation. The problem is that the stars in the 
Mount Wilson survey are relatively faint \citep[$V \sim$ 
6;][]{1995ApJ...438..269B}, making them unfit for dedicated ground-based 
asteroseismic networks like {\it SONG} \citep{2008JPhCS.118a2041G}, which are 
needed to obtain the extremely high frequency precision required for the 
analysis. The reason for this is of course that bright $G$ and $K$ stars 
are rare in the sky, the few exceptions are: $\alpha$ Cen A \& B ($V$ = 
--0.01 and 1.33, respectively), 70 Oph A ($V$ = 4.03), $\kappa^1$ Cet ($V$ = 
4.83) and the solar twin 18 Sco ($V$ = 5.5). It is therefore clear that 
the {\it Kepler} mission offers a unique possibility for sounding stellar 
cycles.

Once we have obtained a firm understanding of the solar cycle, we might be 
able to use the fact that the Sun seems to have a more complicated dynamo 
than most other solar-type stars as an advantage -- as this may enable 
us to test more complicated dynamo models on the Sun. But the prolonged 
delay in the onset of cycle 24 has clearly shown us that we do not have a 
firm understanding of the solar cycle \citep{zimmerman}.

This paper is the first in a series dedicated to the {\it Kepler} mission's 
study of stellar cycles in solar-type stars. The paper evaluates the 
expected frequency precision for {\it Kepler}, the changes in the behavior of 
the stellar cycles on the main sequence and the effect of the convection 
zone depth and differential rotation on the acoustic spectra. Although the 
analysis in this paper is dedicated to the {\it Kepler} mission, it can also 
serve as a guide for selecting targets for other asteroseismic surveys, 
such as {\it SONG}.

We continue the analysis of the potential for sounding activity cycles in 
solar-type stars by \citet{2007MNRAS.377...17C}, but we now have the 
possibility to use specific estimates of the frequency precision for the 
{\it Kepler} mission. We also discuss problems with the way that the acoustic 
amplitude of the stellar cycles was estimated by 
\citet{2007MNRAS.377...17C}. The analysis of the signature of stellar 
rotation in the acoustic spectrum follows \citet{2008A&A...485..813C}, but 
here we extend the analysis to include (1) updated estimates of mode lifetimes and the effect of these on the possibility of detecting the small rotational frequency splittings, and (2) the effect of differential rotation. Part of the 
analysis in this paper was done within asteroFLAG hare-and-hounds 
exercises as part of the preparation for the {\it Kepler} mission 
\citep{2008AN....329..549C} and some of the new scaling relations presented in this paper have thus been obtained from and tested on the artificial stars 
generated as part of the AsteroFLAG hare-and-hounds exercises.

The layout of the rest of the paper is as follows. In section~2 we review 
the details of the study of solar-type stars under the {\it Kepler} 
Asteroseismic Investigation (KAI). A review of the planned ground-based 
support for the study of stellar cycles in solar-type stars under the KAI is given in 
section~3 and an analysis of the expected frequency precision is given in 
section~4. Sections~5--7 use a number of isochrone tracks to estimate the 
effect of rotation (section~5), stellar cycles (section~6) and the 
convection zone (section~7) on the oscillation mode frequencies that we 
expect to measure from the observations. This analysis is used in a 
discussion of the best strategy for selecting the optimal full-length of 
the mission {\it Kepler} targets to increase our understanding of stellar cycles 
(section~8). Concluding remarks are found in section~9.

\section{Kepler observations}

\begin{figure}
\includegraphics[width=\columnwidth]{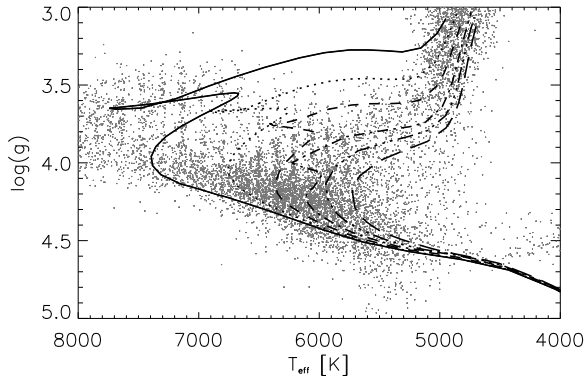}
\caption{Padova isochrone tracks used in the analysis, shown as a function 
of effective temperature and surface gravity. The different line styles 
give the age of the stars: the solid line represents stars that are 1.0 
Gyr old, dotted 1.6 Gyr, dashed 2.5 Gyr, dot-dashed 4.0 Gyr, 
dot-dot-dot-dashed 6.3 Gyr and long dashed 10.0 Gyr. This notation is used 
for the rest of the paper. The gray dots are stars from the {\it Kepler Input 
Catalog} with $g$-band magnitudes between 8 and 10.} 
\end{figure}

{\it Kepler} is a NASA Discovery class space mission that was launched into an 
Earth trailing orbit on 7 March 2009, with the primary goal of detecting 
terrestrial planets using the transit method \citep{2005NewAR..49..478B}. 
The planned mission lifetime of {\it Kepler} is 42 months (excluding commissioning) with an optional extension of 30 months. The satellite is expected to be operational beyond the extended mission, but the downlink capability will be limited due to the distance between the satellite and the Earth. {\it Kepler} will monitor over 100,000 stars with a cadence of 30 min. In parallel with the planet program, {\it Kepler} also has an Asteroseismic Investigation 
\citep[KAI;][]{2007CoAst.150..350C}.

The KAI will begin with an initial run of 8 -- 10 months depending on the length of the commission phase. In this survey phase 
of the mission, targets will only be observed for one month each. In the 
survey phase, close to 40,000 pixels are available for the KAI in short cadence (60 
s). After the survey phase the KAI will shift to a specific target 
phase for the rest of the mission, and it is among these targets that it 
will be possible to study stellar cycles. In the specific target phase, 
the KAI will have the possibility to allocate 11,800 pixels for 
observations in short cadence. A conservative estimate is that around 3,000 
of these 11,800 pixels will be allocated to $F$, $G$, and $K$ stars on the 
main sequence suitable for studying stellar cycles. The reason for giving 
these numbers in pixels and not as a quantity of targets is that bright 
stars will require more pixels than faint stars due to saturation. It is 
estimated that a 6th magnitude star will require 2,640 pixels, whereas a 
12th magnitude star will only need 70 pixels. In practice, this means 
that the specific targets for studying stellar cycles could be chosen as 
e.g.: one 6th magnitude, ten 8th magnitude or 43 13th magnitude stars. The 
numbers of pixels that need to be allocated for stars at a given brightness 
are likely to decrease after the survey phase, when the characteristics of the 
instrument are better known. {\it Kepler} is designed to provide nearly photon 
noise limited photometry on saturated stars. The concept of doing 
photometry on saturated stars has been demonstrated with 
HST observations \citep{2008AJ....136..566G}.

The targets for the survey phase have been chosen from the {\it Kepler Input 
Catalog} \citep{2005AAS...20711013L}, and specific targets to be observed for the full-length of the mission must be chosen from the targets in the survey. It is expected that 
only observations from the first three months with be available before the 
specific targets need to be chosen, so in practice only targets from the 
first three months of observations can be selected as specific full-length 
of the mission targets. Fig.~1 shows all of the stars in the {\it Kepler Input 
Catalog} with $g$-band magnitude between 8 and 10, with Padova isochrone 
tracks \citep{2004A&A...415..571B, 2002A&A...391..195G, 
2004A&A...422..205G} plotted on top as a function of effective temperature 
and surface gravity. The isochrone tracks were calculated for 6 different 
ages between 1 and 10 Gyr in steps of 0.2 dex, using a metallicity of 
$z$ = 0.02. The same stars are plotted in Fig.~2, but here the targets selected 
for observations in the first three months of the survey phase are marked. 
It is seen that $F$ stars are highly over-represented compared to cooler 
stars in the first three months of the survey phase and in general. 
Although this reflects the stellar populations at a given magnitude bin, 
it is unfortunate as $F$ stars are the ones that are hardest to analyze 
with asteroseismology since they do not obey the asymptotic frequency relation \citep{1980ApJS...43..469T} and have short mode lifetimes \citep{chaplin2009}. The fraction of stars which show stellar cycles 
is also significantly larger among $G$ and $K$ stars than among $F$ stars 
\citep{1995ApJ...438..269B}.

\begin{figure}
\includegraphics[width=\columnwidth]{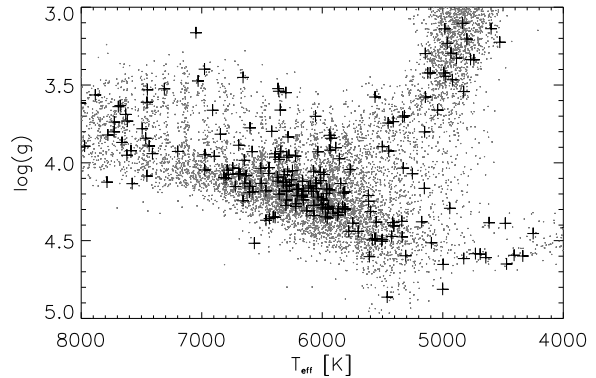}
\caption{All stars in the {\it Kepler Input Catalog} with $g$-band magnitudes 
between 8 and 10 (gray dots) and targets chosen for observations in the 
first three months of the survey phase (black crosses).}
\end{figure}

\section{Ground-based observations}

Complementary ground-based chromospheric observations are important for 
mainly two reasons: firstly, they can 
be used to compare the cycle-induced changes in the chromosphere to 
changes in the amplitudes and frequencies of the oscillation modes -- in 
other words, the complementary data provide a possibility to link the 
changes near the surfaces of the stars (i.e. in the chromospheres) to 
changes in the stellar interiors (i.e. in the oscillation modes). 
Secondly, ground-based chromospheric observations will be important if the 
cycles are significantly longer than the {\it Kepler} mission lifetime, or if 
the cycle-induced changes in the mode frequencies and amplitudes are too 
small to be measured with seismology. We will analyze the expected cycle 
lengths and seismic amplitudes in section~6.

We plan to measure the chromospheric activity of the coolest 
candidate targets discussed in section~2 during the summer of 2009 using 
the high-resolution FIbre-fed Echelle Spectrograph (FIES) mounted on the 
2.6 meter Nordic Optical Telescope \citep{2000mons.proc..163F}. After the 
specific targets have been selected, we will start to monitor the targets 
to look for stellar cycles in those stars. Note that even stars which show 
no sign of stellar cycles will be interesting, since it might be possible 
to use the asteroseismic measurement of the depths of the convection zones 
and rotation to understand why these stars show no sign of stellar cycles.

\section{Expected frequency precision}

\begin{figure}
\includegraphics[width=\columnwidth]{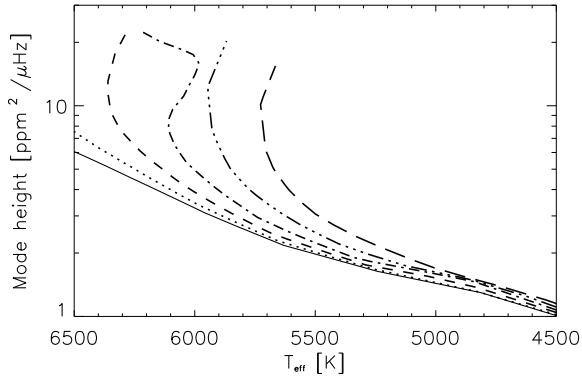}
\caption{Mode height as a function of effective temperature for the Padova 
isochrone tracks. It is seen that the mode height increases dramatically 
when the stars begin to evolve off the main sequence.}
\end{figure}

To estimate the potential for sounding stellar cycles, and for measuring 
the depths of the convection zones and differential rotation in the stars 
observed by {\it Kepler}, we first need to know what frequency precision we will 
be able to obtain from the observations. The frequency precision is 
determined by three parameters: the S/N of the oscillation modes in the 
acoustic spectrum, the mode lifetime $\tau$ and the length of the 
observations $T$ as given by the formulation of 
\citet{1992ApJ...387..712L}:
\begin{equation}
\sigma_{\nu}=\sqrt{\frac{F(\beta)\Delta}{4\pi T}},
\end{equation}
where $F(\beta)$ is a function of the inverse S/N, $\beta$:
\begin{equation}
F(\beta)=\sqrt{(1+\beta)}\left(\sqrt{(1+\beta)}+\sqrt{\beta}\right)^3,
\end{equation}
and $\Delta = 1/(\pi \tau)$ is the mode FWHM linewidth. The Libbrecht 
formulation provides a lower limit on the frequency precision, since 
it does not account for blending effects in closely spaced oscillation 
modes. This is mainly a problem for modes with different azimuthal order, 
which is analyzed in detail in section~5.

We use the $T_{\rm eff}^{-4}$ scaling relation from \citet{chaplin2009} 
for the mode lifetimes, which also provides us with the following relation 
for the height $H$ of the oscillation modes in the acoustic power density spectrum by using the scaling relation for the amplitude 
from \citet{1995A&A...293...87K} for intensity observations:
\begin{equation}
H \propto g^{-2},
\end{equation}
where $g$ is the surface gravity. Using the Padova isochrone tracks, we 
have plotted the mode height as a function of effective temperature in 
Fig.~3. It is seen that the mode height increases dramatically when the 
stars begin to evolve off the main sequence. It is also seen that the mode 
height decreases toward lower temperature, from 3 ${\rm ppm}^2/\mu{\rm 
Hz}$ for the Sun down to just over 1 ${\rm ppm}^2/\mu{\rm Hz}$ for stars 
with temperatures around 4500 K.

\begin{figure}
\includegraphics[width=\columnwidth]{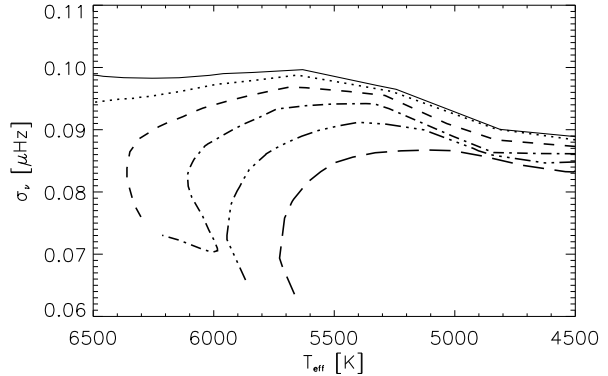}
\caption{Frequency precision after 1 year of observations of a 9th 
magnitude star with {\it Kepler} as a function of effective temperature for the 
Padova isochrone tracks. The frequency precision scales with $1/\sqrt{T}$ 
where $T$ is the length of the observations. The frequency precision after 
4 years of observations can therefore be calculated by dividing the 
numbers on the y-axis by 2.}
\end{figure}

Knowing the expected mode heights, we can use the expected photometric 
errors of the {\it Kepler} photometer to calculate the expected S/N. The 
photometric error is a function of many factors -- the main one being the 
magnitude. It is estimated that the photometric error for a 60 sec 
integration $\sigma_{\rm int}$ with {\it Kepler} will be 45 ppm at 8th 
magnitude, 69 ppm at 9th magnitude and 111 ppm for a 10th magnitude star. 
The photometric errors are based on laboratory tests with fully flight-like electronics with an assumed level of stellar (activity and granulation) noise of 10 ppm \citep[see][for details of the laboratory tests]{2000SPIE.4013..508K}. 
Assuming that the noise is random and uncorrelated, the photometric error 
can be converted into a power density error $\sigma_{\rm pd}$ using:
\begin{equation}
\sigma_{\rm pd} = 2T_{\rm int} \sigma_{\rm int}^2,
\end{equation}
where $T_{\rm int}$ is the sampling time -- i.e. 60 sec. In this way, 
we expect the error in the power density spectrum to be 0.24, 0.57 and 
1.48 ${\rm ppm}^2/\mu{\rm Hz}$ for a 8th, 9th and 10th magnitude star, 
respectively. In comparison {\it MOST}, {\it WIRE} and {\it CoRoT} have obtained noise levels in the high-frequency part of the power density spectrum near the Nyquist frequency as low as 3, 1, and 0.15 ${\rm ppm}^2/\mu{\rm Hz}$, respectively \citep{2008ApJ...687.1448G, 2005ApJ...633..440B, 2008A&A...488..705A}. These values are not directly comparable to the expected error in the power density spectrum that we calculate for $Kepler$ since they are measured at high frequency and thus include very little activity and granulation signal, but it is clear that the expected error that we calculate does not seem unrealistic in comparison.

The estimated frequency precision for a 9th magnitude star as a function 
of effective temperature is shown in Fig.~4. It is seen that the evolved 
stars will yield the best frequency precision. It is also seen that the 
frequency precision improves toward lower effective temperature, from 
around the effective temperature of the Sun down to an effective 
temperature of 4500 K, but the slope is small. In other words, although 
the mode height changes on the main sequence, the frequency precision for 
oscillation modes in stars on the main sequence is approximately constant.

\section{Prediction of rotation}

\begin{figure}
\includegraphics[width=\columnwidth]{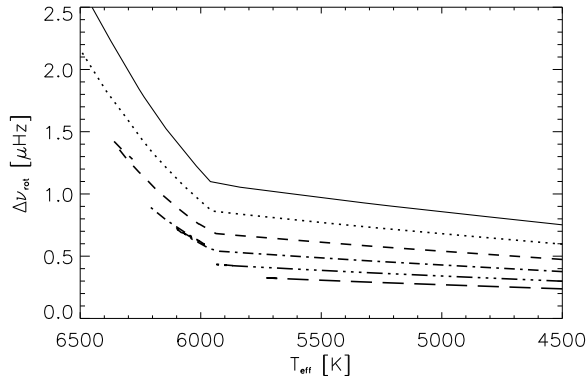}
\caption{Rotation frequency splitting as a function of effective 
temperature for the Padova isochrone tracks. It is seen that the 
rotational splitting decreases toward lower effective temperature.}
\end{figure}

The mean surface rotation period $P_{\rm rot}$ can be estimated from 
the empirical relationship given by \citet{2004A&A...414.1139A}. This 
relationship is derived from photometric observations of stars in the 
Hyades \citep{1987ApJ...321..459R, 1995ApJ...452..332R}. The relationship 
is:
\begin{equation}
\begin{array}{l} 
\mathrm{log}(P_{\rm rot}) - 0.5 {\rm log}(t/0.625) = \\
\left\{\begin{array}{rll} 
-0.669 + 2.580 (B - V) & {\rm for}  & 0.45 \leq B -V \leq 0.62\\
0.725 + 0.326 (B - V)   & {\rm for}  & 0.62 < B - V \leq 1.30
\end{array} \right.
\end{array}
\end{equation}
Here, $t$ is the age of the star in Gyr, and it appears in the 
relationship because the $t^{1/2}$ spin-down law of 
\citet{1972ApJ...171..565S} is implicit. 

When measuring stellar rotation with asteroseismology, it is the rotational frequency splitting $\Delta \nu_{\rm rot}$ that is measured rather than the rotation period. Assuming that the stars possess internal rotation rates that are comparable to the surface rates, as in the Sun, it is reasonable to assume that the mean rotation rate is given as the inverse of the mean rotation period. This might be questionable for younger stars \citep[see][for discussion]{2008A&A...485..813C}. As the oscillations that we are studying in this paper are high-order p modes we can also assume that the rotation frequency splittings equal the rotation rates. Note that this is not the case for, e.g. the high-order g modes in white dwarfs where the rotational splitting is only equal to half the rotation rate \citep{JCD}. Under these assumptions the rotation period can easily be converted into an equivalent rotational frequency splitting by using the relation: $P_{\rm rot} = 1/\Delta \nu_{\rm rot}$. 

Fig.~5 shows the rotational frequency splitting as a function of effective 
temperature for the Padova isochrone tracks. It is generally seen that the 
rotational splitting deceases toward lower effective temperature and that 
the slope of this trend is much larger for stars hotter than the Sun 
compared to stars cooler than the Sun.

\begin{figure}
\includegraphics[width=\columnwidth]{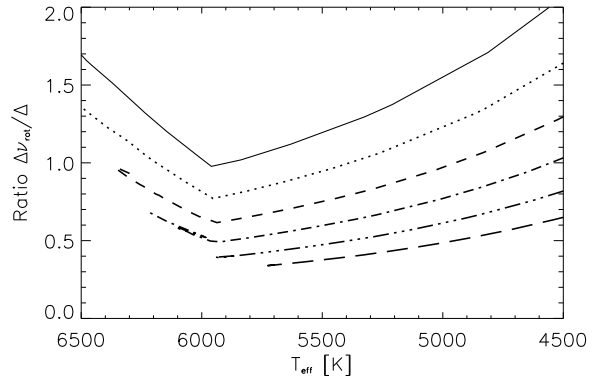}
\caption{Ratio between rotational frequency splitting and the mode 
linewidth as a function of effective temperature for the Padova isochrone 
tracks. Note the interesting conclusion that the effective temperature of 
the Sun is the worst possible effective temperature at which rotational 
frequency splittings can be measured. This provides some hope that we will 
be able to measure differential rotation with asteroseismology.}
\end{figure}

\subsection{Effect of finite mode lifetime}
The possibility of measuring the rotational frequency splitting, and thus 
the mean rotation period of the stars, relies not only on a frequency 
precision sufficient to measure the small splittings, but also on 
splittings that are relatively large compared to the mode linewidth. If 
the linewidth is significantly larger than the splitting, it will not be 
possible to disentangle the modes of different azimuthal order $m$, and it 
will thus not be possible to measure the splittings. For this reason the internal rotation rate of the Sun was first obtained from low-degree helioseismology when it was realized that only low-frequency modes with long mode lifetime should be used for this analysis \citep{1995Natur.376..669E}. Fig.~6 shows the 
ratio between the rotational frequency splittings and the mode linewidth, 
which can serve as a measure of the chances of disentangling the modes 
with different azimuthal order. The figure shows the surprising result 
that the effective temperature of the Sun is the worst possible effective 
temperature at which rotational frequency splittings can be measured. 

As shown by \citet{2004SoPh..220..269G}, it is just possible to measure 
radial differential rotation in the low-order frequency splittings from 6 
years of low-degree observations of the Sun. Since the ratio between the 
rotational frequency splitting and the mode linewidth will be larger for 
all other stars on the main sequence, it is therefore not unreasonable to 
hope that we will be able to use {\it Kepler} to measure radial differential 
rotation in some solar-type stars with rotation profiles similar to the Sun.

In order to measure stellar rotation the rotation axis of the star needs to be inclined with respect to the line-of-sight. If the inclination angle between the rotation axis of the star and the line-of-sight is low, then the rotation will not significantly affect the frequencies of the observed oscillation modes with different azimuthal order, and the rotational frequency splitting can thus not be measured. \citet{2003ApJ...589.1009G} find that it is only possible to measure the rotational frequency splitting if the inclination angle is larger than around 30$^{\circ}$. 
Statistically, assuming random orientation of the rotation axis, it will therefore not be possible to measure the rotational frequency splitting in around 15 per cent of the targets due to low inclination angles.

\subsection{Effect of differential rotation}
The Sun rotates fastest in the outer part of the convection zone beneath 
the equator, with a rotation rate of around 470 nHz. The rotation rate 
decreases to around 435 nHz at the base of the convection zone, and to 
around 450 nHz at the surface -- with a further decrease along the surface 
to the poles with a value of around 370 nHz at 60$^{\circ}$ which is the highest latitude at which rotation rates can be measured reliably with helioseismology \citep{1998ApJ...505..390S}. The largest gradients in the rotation rate are generally seen near the surface and near the base of the convection zone.

Differential rotation in the Sun and in other solar-type stars will 
manifest itself both as latitudinal and radial differential rotation. 
Latitudinal differential rotation is mainly on the surface, and can be 
measured either with asteroseismology or by fitting a spot model to the 
observed intensity time series as has been done for the stars $\kappa^1$ 
Ceti and $\epsilon$ Eri using the $MOST$ satellite 
\citep{2006ApJ...648..607C, 2007ApJ...659.1611W}. Radial differential 
rotation is confined to the stellar interior, and can only be measured 
with asteroseismology.

Differential rotation can be measured with asteroseismology by measuring 
differences in the separation between modes with different azimuthal order 
$m$ (the rotation splitting) over different radial orders $n$ and angular 
degrees $l$. A simplified way of doing this is by averaging the rotational 
splittings over different radial orders, and thus obtaining the splitting 
only as a function of the angular degree 
\citep[see][]{2003ApJ...589.1009G, 2004SoPh..220..169G, 
2006MNRAS.369.1281B}. This is done to increase the S/N of the measured 
splittings, but the approach in strongly biased towards measuring 
latitudinal differential rotation on the surface, and is much less 
efficient at measuring radial differential rotation at the base of the 
convection zone. For the Sun, the mean difference in the rotational 
splitting between $l=1$ and $l=2$ modes is around 7 nHz at low frequency 
\citep{2003ESASP.517..271G}. We therefore adopt this as a conservative 
reference value, although there might be better ways of averaging the 
rotational splittings for stars with thicker convection zones and thus 
larger radial rotation gradients \citep{2007ApJ...657..486B}. Such new 
averaging schemes would need to be calculated based on detailed forward 
modeling of the expected changes in the rotational splitting, not only 
over different angular degrees, but also over different radial orders. In 
this way, it would hopefully be possible to measure average rotational 
splittings that would be more sensitive to radial changes in the rotation 
rate at the base of the convection zone than to latitudinal changes in the 
rotation rate at the surface.

\begin{figure}
\includegraphics[width=\columnwidth]{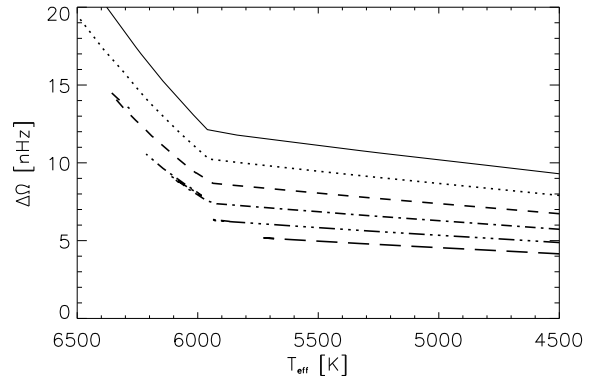}
\caption{The difference in the rotational splitting between $l=1$ and 
$l=2$ modes averaged over different radial orders as a function of 
effective temperature for the Padova isochrone tracks. It is seen that the 
differences increase toward higher temperature and decrease with age.}
\end{figure}

Since only surface (latitudinal) differential rotation has been measured 
in other solar-type stars so far, we are only able to make predictions of 
the changes in the surface differential rotation on the main sequence. 
Surface differential rotation is normally expressed as the change in the 
rotation period $\Delta P_{\rm rot}$ over the stellar disc. A relation between 
$\Delta P_{\rm rot}$ and $P_{\rm rot}$ was obtained by 
\citet{1996ApJ...466..384D} by looking at the changes in the rotation 
period measured in at least five different observing seasons in the 
stellar Ca~{\sc ii} measurements from Mount Wilson Observatory:
\begin{equation}
\Delta P_{\rm rot} \propto P_{\rm rot}^{1.3 \pm 0.1}.
\end{equation}
The interesting parameter for asteroseismology is not the change in the 
rotation {\em period} $\Delta P_{\rm rot}$ caused by differential 
rotation, but rather the change in the rotation {\em rate} $\Delta 
\Omega$. This can easily be obtained by assuming that $\Delta P_{\rm rot} 
\ll P_{\rm rot}$ and thus $\Delta \Omega \propto \Delta P_{\rm rot} / 
P_{\rm rot}^2$, which gives the following relation between the changes in 
the rotation rate caused by differential rotation and the mean rotation 
period:
\begin{equation}
\Delta \Omega \propto P_{\rm rot}^{-0.7 \pm 0.1}.
\end{equation}

Adopting 7 nHz as the mean difference in the rotational splitting between 
$l=1$ and $l=2$ modes at low frequency \citep{2003ESASP.517..271G} in the 
Sun, we can plot the mean difference shown in Fig.~7. It is seen that the 
differences increase toward higher temperature and decrease with age.

The result in Fig.~7 should be seen as a lower limit on the effects of 
differential rotation that can be detected with asteroseismology, both 
because there might be better ways to measure radial differential rotation 
(as discussed above) and because differential rotation measured through 
changes in the rotation period samples a limited range of latitudes 
\citep[see][for discussion]{1996ApJ...466..384D}.

Slightly higher values of the exponent in Eq.~6 have been found by 
\citet{1995ApJS...97..513H}, who obtained an exponent of $1.76 \pm 0.06$ 
based on photometry of active binaries, and \citet{2005MNRAS.357L...1B} 
who obtained $1.85 \pm 0.10$ based on Doppler imaging techniques and 
spectral analysis \citep{2003A&A...398..647R}. We have chosen to use the 
relation by \citet{1996ApJ...466..384D} because the sample used in our 
study is not biased towards really active stars (as the photometry 
results) or towards really fast rotating stars (as the Doppler imaging and 
the spectral analysis results). The effect of increasing the exponent in 
Eq.~6 would be to flatten the curves in Fig.~7.

Differential rotation in solar-type stars has also been studied with 
mean-field models by \citet{2008JPhCS.118a2029K}, who see a similar 
relation between $\Delta \Omega$ and the effective temperature as we see 
in Fig.~7 -- i.e. the slope of the curves seems to increase dramatically 
at an effective temperature near 6000 K.

\section{Prediction of stellar cycles}

\begin{figure}
\includegraphics[width=\columnwidth]{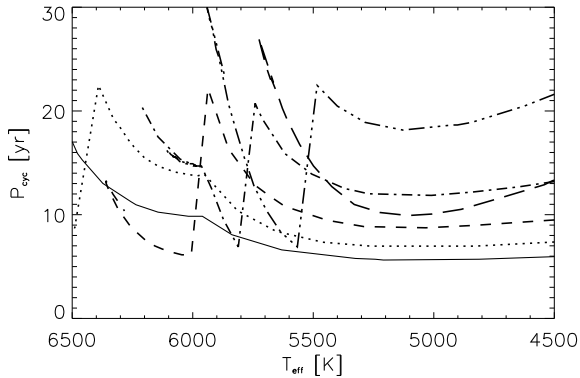}
\caption{Stellar cycle period as a function of effective temperature for 
the Padova isochrone tracks. No general trend is seen.}
\end{figure}

A commonly used indicator of surface activity on stars is the Ca~{\sc ii} 
H and K emission index. This index is usually expressed as $R'_{\rm HK}$ 
-- the average fraction of the stellar luminosity that is emitted in the 
Ca~{\sc ii} H and K emission line cores \citep[see][for 
discussion]{2007AJ....133..862H}. Using the first data from the Mount 
Wilson survey, \citet{1984ApJ...279..763N} were able to obtain relations 
between the rotation period $P_{\rm rot}$, the $R'_{\rm HK}$ index and the 
$B-V$ color, which can be used to obtain a scaling relation for the 
$R'_{\rm HK}$ index:
\begin{equation}
{\rm log}(\tau_c / P_{\rm rot}) =  - (0.324 -  0.400y + 0.283y^2 - 1.325y^3), 
\end{equation}
where $y = $log$(R'_{\rm HK} \times 10^5)$ and $\tau_c$ is the convective 
turnover time. The convective turnover time may then be calculated from:
\begin{equation} 
\begin{array}{l} 
{\rm log}(\tau_c)  = \\ \left\{ \begin{array}{rrll} 
1.361 + 0.166x & + 0.025x^2 - 5.323x^3 & {\rm for}  & x \geq 0\\
1.361 - 0.140x   &  &{\rm for}  & x < 0
\end{array} \right.
\end{array}
\end{equation}
where $x = 1 - (B - V )$ and the convective turnover time is measured in 
days. To calculate the $R'_{\rm HK}$ index, we solve the cubic Eq.~8 using 
the known input parameters: $B - V$ and $P_{\rm rot}$.

Knowing the $R'_{\rm HK}$ index, we can calculate the cycle period $P_{\rm 
cyc}$ from the relation given by \citet{2008A&A...485..813C}:
\begin{equation}
\begin{array}{l} 
{\rm log} (P_{\rm cyc}) = \\ \left\{ 
\begin{array}{rll} 
- 6.7 - 1.6 {\rm log} (R'_{\rm HK}) & {\rm for}  & {\rm log} (R'_{\rm HK}) < - 4.75\\
- 5.5 - 1.4 {\rm log} (R'_{\rm HK}) & {\rm for}  & {\rm log} (R'_{\rm HK}) \geq - 4.75\\
\end{array} \right.
\end{array}
\end{equation}

Fig.~8 shows the expected stellar cycle period as a function of effective 
temperature. No general trend is seen. For the three youngest isochrones, 
it is seen that the cycle period generally increases toward higher 
temperature. For effective temperatures lower than the effective 
temperature of the Sun, it is seen that the cycle periods generally 
increase with age as the stars evolve from active to inactive stars while 
their rotation periods decrease. On the other hand, as the stars evolve 
from the active to the inactive sequence, the number of rotations in each 
stellar cycle decreases and therefore no general trend is seen in the 
figure.

\begin{figure}
\includegraphics[width=\columnwidth]{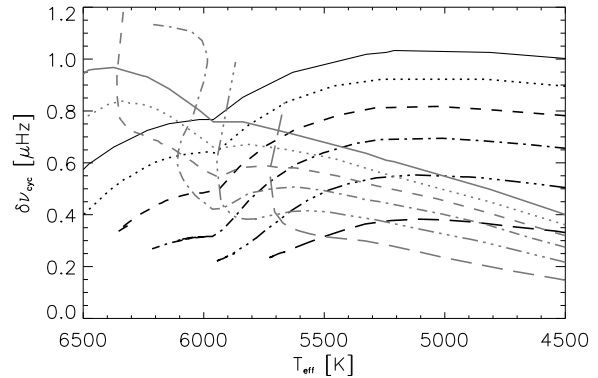}
\caption{Stellar cycle frequency shifts as a function of the effective 
temperature for the Padova isochrone tracks. For the black lines the 
shifts were calculated using the scaling relation $\delta \nu_{\rm cyc} 
\propto \Delta R'_{\rm HK}$ from \citet{2007MNRAS.377...17C} and for the 
gray lines the shifts were calculated using the scaling relation $\delta 
\nu_{\rm cyc} \propto \frac{R^{2.5}L^{0.25}}{M^2} \Delta R'_{\rm HK}$ from 
\citet{2007MNRAS.379L..16M}. It is seen that the shifts calculated using 
the scaling relation by \citet{2007MNRAS.377...17C} are nearly constant 
for stars cooler than 5500 K, while at the same age the shifts calculated 
using the scaling relation by \citet{2007MNRAS.379L..16M} generally 
decrease around 50 per cent between 5500 K and 4500 K.}
\end{figure}

The amplitudes of the activity cycles $\Delta R'_{\rm HK}$ can be obtained 
from the relation given by Saar \& Brandenburg (2002): 
\begin{equation} 
\Delta R'_{\rm HK} \propto ( R'_{{\rm HK}} )^{0.77}. 
\end{equation}
The p-mode frequencies in the Sun are generally believed to be affected by 
the solar cycle through changes in the turbulent velocity in the outermost 
regions of the near surface convection zone, caused by changes in the RMS 
magnetic field in the Sun \citep{2004ApJ...600..464D, 
2005ApJ...625..548D}. Relations between the Ca~{\sc ii} amplitudes of the 
activity cycles $\Delta R'_{\rm HK}$ and the cycle frequency shifts 
$\delta \nu_{\rm cyc}$ have been proposed by \citet{2007MNRAS.377...17C} 
and \citet{2007MNRAS.379L..16M}.

\citet{2007MNRAS.377...17C} proposed that the cycle frequency shifts 
simply scale linearly with the amplitudes of the activity cycles:
\begin{equation}
\delta \nu_{\rm cyc} \propto \Delta R'_{\rm HK}.
\end{equation}
Evidence for this relation is found from observations of low-degree 
oscillations in the Sun, where it is seen that the cycle frequency 
shifts change linearly with changes in the Mg~{\sc ii} lines over the 
solar cycle \citep{2007ApJ...659.1749C}. The expected stellar cycle 
frequency shifts as a function of the effective temperature obtained using 
the scaling from \citet{2007MNRAS.377...17C} is shown in Fig.~9 (black 
lines). It is seen that the cycle frequency shifts generally increase 
toward lower temperature and decrease with age.

On the other hand, \citet{2007MNRAS.379L..16M} assume that the cycle 
frequency shifts scale as
\begin{equation} 
\delta \nu_{\rm cyc} \propto \frac{D}{I}\Delta R'_{\rm HK},
\end{equation} 
where $D$ is the depth of the source of the perturbations beneath the 
photosphere and $I$ is the mode inertia. The inverse relation with the 
mode inertia is known from studies of solar cycle mode frequency shifts 
\citep[see][for discussion]{1990Natur.345..779L, 2004ApJ...600..464D, 
2005ApJ...625..548D}. \citet{2007MNRAS.379L..16M} then assume that the 
mode inertia scales as $M/R$ and that the source depth scales with the 
pressure scale height at the photosphere $H_p$, which itself scales as 
$L^{0.25}R^{1.5}/M$. We thus obtain:
\begin{equation} 
\delta \nu_{\rm cyc} \propto \frac{R^{2.5}L^{0.25}}{M^2} \Delta R'_{\rm HK},
\end{equation} 
Note that we lack a physical argument for the relation between the source 
depth and the pressure scale height other than the fact that most things 
in the near surface layers scale with the pressure scale height. The {\it Kepler} observations might eventually provide us with such arguments as it might be possible to extract information of the location of the source of the p-mode excitation from the high-frequency oscillations in the solar-type stars as has been done for the Sun \citep{1991ApJ...375L..35K} and other solar-type stars \citep{2007MNRAS.381.1001K}, but it remains to be proven if similar kinds of analysis could also reveal the location of the source of the perturbations caused by the stellar cycles.

Fig.~9 also shows the expected stellar cycle frequency shifts as a 
function of the effective temperature obtained by using the scaling from 
\citet{2007MNRAS.379L..16M} (gray lines). It is seen that the shifts 
generally increase toward higher temperature and decrease with age. The 
shifts calculated using the scaling relation by 
\citet{2007MNRAS.377...17C} are nearly constant for stars cooler than 5500 
K but with the same age. On the other hand, the shifts calculated using 
the scaling relation by \citet{2007MNRAS.379L..16M} generally decrease 
around 50 per cent between 5500 K and 4500 K for a given age. It is currently not 
possible to determine which of the two scaling relations is more 
appropriate, but it is clear from Figs.~4~\&~9 that we will be able to measure 
the oscillation mode frequencies precisely enough from one year of {\it Kepler} 
data to detect stellar cycle frequency shifts when these frequencies are 
compared to {\it Kepler} observations obtained in other years. We also note that 
the {\it Kepler} observations might allow us to test which of the scaling 
relations is the better one.
\section{Prediction of the convection zone depth}

One of the most important things that will help us understand the stellar 
cycles we observe with {\it Kepler} is a measurement of the convection zone 
depth. The depth of the convection zone can be measured with 
asteroseismology by analyzing the second differences in the frequencies:
\begin{equation}
\Delta_2 \nu_{n,l} \equiv \nu_{n-1,l}-2 \nu_{n,l}+ \nu_{n+1,l}
\end{equation}
Localized regions of rapid variation of the sound speed, such as the 
bottom of the convection zone, will induce an oscillatory component in the 
second frequency differences as a function of frequency 
\citep{2000MNRAS.316..165M, 2007MNRAS.375..861H}. Other localized regions 
that cause oscillatory components in the second frequency differences are 
the ionization zones of He~{\sc i} and He~{\sc ii}. 
\citet{2007MNRAS.375..861H} have given full analytical solutions for how 
to represent the oscillatory components in the second frequency 
differences from the three regions. For the analysis here, we are only 
interested in determining how the amplitude of the oscillatory component 
originating from the base of the convection zone depends on fundamental 
stellar parameters.

To evaluate how the amplitude of the oscillatory component 
originating from the base of the convection zone changes with fundamental 
stellar parameters, we have used frequencies from 6 artificial solar-type 
stars generated as part of the asteroFLAG hare-and-hounds exercises 
\citep{2008AN....329..549C}. The amplitudes of the oscillatory components 
in these stars were measured by calculating the least-squares periodogram 
of the second differences. These periodograms all showed 3 peaks from the 
3 regions of rapid variation of the sound speed, and the peak with the 
highest frequency is the one that originates from the base of the 
convection zone. We then looked for correlations between the amplitude of 
the peak originating from the base of the convection zone and the 
effective temperature and surface gravity. It was seen that there was a 
clear correlation between the logarithm of the surface gravity ${\rm 
log}(g)$ and the amplitude of the form:
\begin{equation}
A_{\rm cz} \propto {\rm log}(g).
\end{equation}
We adopt a value of 0.4 $\mu$Hz for the mean amplitude of the oscillatory 
component in the second frequency differences in the Sun 
\citep{2007MNRAS.375..861H}. Fig.~10 shows the amplitude of the 
oscillatory component originating from the base of the convection zone as 
a function of the effective temperature. It is seen that the amplitude 
generally increases toward lower temperature.

\begin{figure}
\includegraphics[width=\columnwidth]{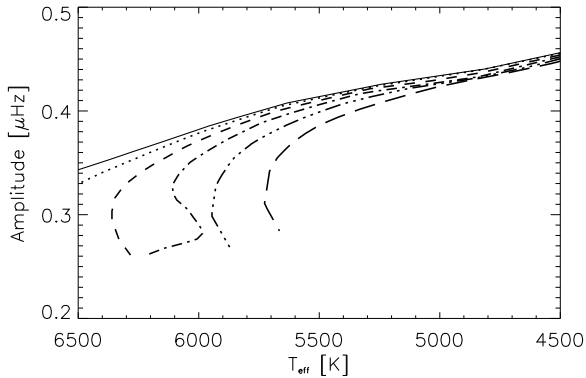}
\caption{The amplitude of the oscillatory component in the second 
frequency differences originating from the base of the convection zone as 
a function of the effective temperature for the Padova isochrone tracks.}
\end{figure}
\citet{2009A&A...493..185R} has recently suggested another approach for 
measuring the convection zone depth with asteroseismology. This method 
makes use of the second and third frequency differences between modes of 
angular degree 0 and 1 instead of the second differences of modes with the 
same degree. This method has the advantage that the amplitude of the 
oscillatory component of the differences is constant as a function of 
frequency -- whereas it decreases toward higher frequencies for the 
differences from modes with the same degree. Also, the method can be tuned 
so that the signal from the He~{\sc ii} ionization zone cancels out, which 
makes it easier to detect the signal from the base of the convection zone. 
However, the method has two weaknesses compared to the differences from 
modes with the same degree. Firstly, it only uses information from modes 
with degree 0 and 1, and not from modes with degree 2 that we will also 
detect in the {\it Kepler} observations (which means that not all information 
from the observations is used to measure the depth of the convection 
zone). Secondly, it is not clear whether the signal from the He~{\sc ii} 
ionization zone can be canceled out in stars other than the Sun.

The signal from the base of the convection zone in the method by 
\citet{2009A&A...493..185R} will also scale with log($g$) as suggested in 
Eq.~16, though the absolute amplitude and S/N could be different from the 
amplitudes and S/N obtained for the second differences from modes with the 
same degree.
\section{Discussion}

Below we summarize the different recommendations for selecting 
asteroseismic targets for sounding stellar cycles.

\subsection{Effect of stellar brightness and S/N of observations}

At 8th magnitude the S/N of the oscillation modes will allow the detection 
of modes in all stars, and the frequency precision will mainly be limited 
by (1) the mode lifetime, which will cause the power to be smeared out 
over a larger frequency range, and (2) the length of the observations. 9th 
magnitude stars will allow the detection of oscillations in most cases. 
Here the frequency precision for stars with long mode lifetimes will be 
limited not only by the mode lifetime and the length of the observations, 
but also by the S/N. 10th magnitude stars will only allow the detection of 
oscillation modes for temperatures similar to the Sun and hotter, and the 
frequency precision will be limited by the S/N for most mode lifetimes.

\subsection{Optimal stellar parameters for measuring stellar rotation with 
asteroseismology}

The worst place in parameter space for measuring stellar rotation with 
asteroseismology is the place of the Sun -- both hotter and cooler stars 
will have larger rotational splittings relative to the mode linewidths in 
the acoustic spectrum. Differential rotation will generally be measured 
most easily in young stars. For stars hotter than 6000 K, differential 
rotation will be measured most easily in the hottest stars, while there is 
almost no dependence on temperature for stars cooler than 6000 K.

\subsection{Optimal stellar parameters for activity cycles}

The analysis did not provide a clear answer to where on the main sequence 
it will be easiest to sound stellar cycles. It is clear that young stars 
will generally have shorter periods and larger frequency changes caused by 
stellar cycles. In order to use asteroseismology to investigate the 
hypothesis put forward by \citet{2007ApJ...657..486B} -- that the 
Vaughan-Preston gap is caused by two different kinds of dynamos operating 
on either side of the gap --- it is important that stars on either side of 
the gap are observed with {\it Kepler}. The best way to determine which side of 
the gap the stars are on is by determining their ages (see section 8.6).

\subsection{Optimal stellar parameters for measuring the convection zone 
depth}

It will generally be easier to measure the depth of the stellar convection 
zone in cooler stars, since the amplitudes in the second frequency 
differences are expected to be larger than in hotter stars. The analysis 
has shown that we can expect to measure the convection zone depth in all 
KAI full length of the mission specific targets in the analyzed range in effective temperature.

\subsection{General optimal stellar parameters for sounding stellar 
cycles}

From these recommendations it is generally seen that the targets should be 
as young as possible, with the reservation that both sides of the 
Vaughan-Preston gap need to be sampled.

The analysis of the effect of differential rotation and stellar cycles on 
the acoustic spectra favors hotter stars, though the dependence is not 
strong and unambiguous. The effect of the convection zone depth favors 
cooler stars. We thus conclude that stars both cooler and hotter than the 
Sun should be on the KAI full length of the mission specific targets list, 
and to ensure a relatively equal distribution this means that we should 
favor cooler stars since they are clearly under-abundant among the targets 
already selected for observations in the first three months of the survey 
phase.

It is clear that measuring stellar differential rotation will be the most 
demanding task of those analyzed here, and this task will set high demands 
on the frequency precision. This means that it will only be possible to 
measure stellar differential rotation with asteroseismology on {\it Kepler} 
stars after they have been observed for the full length of the mission. On 
the other hand, the analysis has shown that evaluating the potential for 
measuring stellar differential rotation based on the frequency shifts seen 
in the Sun from differential rotation is probably too pessimistic. This 
provides some hope that we will be able to measure stellar differential 
rotation in some of the stars observed with {\it Kepler}.

The analysis has also shown that it is probably going to be possible to 
measure the oscillation mode frequencies precisely enough in one year of 
{\it Kepler} data to be able to see stellar cycle frequency shifts when these 
frequencies are compared to {\it Kepler} observations obtained in other years.  
It also seems possible to measure the depth of the convection zone from 
just one year of {\it Kepler} observations, though the full data set will 
eventually be available for this purpose.
 
It is also worth noting that mode frequency shifts might not be the only 
way to sound the stellar cycles. The stellar cycles will also manifest 
themselves as changes in the amplitudes and lifetimes of the modes 
\citep{2000MNRAS.313...32C}, and possibly also in flare induced 
high-frequency oscillations \citep{2008ApJ...678L..73K}, though the last 
concept remains to be proven.

\subsection{Asteroseismic determination of stellar age}

Since one of the recommendations is that mainly young stars should be 
selected, we need a way to determine the stellar ages. The best tool for 
doing this is asteroseismology, and it is reasonable to assume that a 
rough age estimate can be obtained from the survey observations by 
plotting the small and large frequency separations in the asteroseismic 
HR-diagram \citep[also known as the JC-D diagram, 
][]{2007CoAst.150..350C}. Since we are interested in stars around the 
Vaughan-Preston gap (age 2--3 Gyr), we can safely assume that the lines in 
the asteroseismic HR-diagram are vertical to within the precision needed 
for {\it Kepler} specific target selection. This basically means that we are 
mainly interested in stars which have a small separation equal to the 
small separation in the Sun (6 $\mu$Hz) and larger.

\subsection{Recommendations for {\it Kepler} target selection for sounding 
stellar cycles}

We finally end up with the following selection procedure:
\begin{enumerate}
\item[(1)] Use the targets with magnitudes between 8 and 10 in the first 
3 months of the {\it Kepler} survey phase as candidates.
\item[(2)] Start selecting stars from the cool end.
\item[(3)] Ensure that the oscillations can be seen in the acoustic 
spectrum.
\item[(4)] Ensure that the oscillation modes can be understood in the 
framework of the asymptotic frequency relation.
\item[(5)] Ensure that hints of rotational splitting can be seen.
\item[(6)] Ensure that the small separation is large (6 $\mu$Hz).
\item[(7)] Ensure that both active and inactive stars have been selected 
(based on ground-based observations of chromospheric activity in the stars).
\end{enumerate}

\section{Conclusions}

We have presented an analysis of the potential for using observations from 
the {\it Kepler} Asteroseismic Investigation to improve our understanding of 
solar and stellar cycles. The analysis has shown that it will probably be 
possible to sound stellar cycles with {\it Kepler} in stars brighter than 9th 
magnitude by measuring the changes in the oscillation mode frequencies 
caused by the stellar cycles. It should be possible to measure the depth 
of the convection zone and the mean rotation rate from {\it Kepler} observations 
of stars brighter than 9th magnitude. The analysis has also revealed that 
measuring stellar differential rotation will be highly demanding on the 
frequency precision, and it is clear that it will only be possible after 
the mission has finished. On the other hand, one of the unexpected results 
of the analysis was that the worst place in parameter space for measuring 
stellar rotation with asteroseismology is the place of the Sun -- which 
provides some hope that it might be possible to measure stellar 
differential rotation in some of the stars observed with {\it Kepler}. 

The scaling relations we use for rotation, mode lifetime, activity etc. represent a statistical average of the behavior of many, or several, stars, observed by photometry, spectroscopy and asteroseismology. Provided these relations are accurate, our predictions are therefore those for a notional {\it average} field star. Natural scatter means that e.g. not all stars hotter or cooler than the Sun will present a smaller challenge when it comes to rotation; however, given a sufficient number of stars, we predict that on average, things will be easier.

\section*{Acknowledgments}

We acknowledge the International Space Science Institute (ISSI) and 
European Helio- and Asteroseismology Network (HELAS), a major 
international collaboration funded by the European Commission's Sixth 
Framework Programme. CK acknowledges financial support from the Danish 
Natural Sciences Research Council. TSM acknowledges support from NASA 
grant NNX09AE59G and from the National Center for Atmospheric Research, a 
federally funded research and development center sponsored by the 
U.S.~National Science Foundation. WJC and YE acknowledge the support of 
the UK Science and Technologies Facilities Council (STFC). We thank 
Dr.~Ron Gilliland for providing us with estimates of the expected 
photometric precision for {\it Kepler}.

\label{lastpage}
\end{document}